\documentstyle[aps,multicol,epsfig,psfig]{revtex}
\def\bea{\begin{eqnarray}}
\def\eea{\end{eqnarray}}
\def\be{\begin{equation}}
\def\ee{\end{equation}}
\draft
\begin{document}
\author{Krzysztof Sacha}
\address{
Fachbereich Physik,
Univerisit\"at Marburg,
Renthof 6,
D-35032 Marburg,
Germany\\
 Instytut Fizyki im. Mariana Smoluchowskiego, Uniwersytet
Jagiello\'nski,
ul. Reymonta 4, PL-30-059 Krak\'ow, Poland\\
}
\title{Non-resonant driving of H atom with broken time-reversal symmetry}

\date{\today}

\maketitle{ }

\begin{abstract}
The dynamics of atomic hydrogen placed in a static electric field and
illuminated by elliptically polarized microwaves is studied in the range of
small field amplitudes where perturbation calculations are applicable.
For a general configuration of the fields any generalized time-reversal
symmetry is broken and, as the classical dynamics is chaotic, the
level statistics obeys the random matrices prediction of Gaussian unitary 
ensemble. 
\end{abstract}

\begin{multicols}{2}  
The Bohigas, Giannoni and Schmit conjecture \cite{bohigas84} 
that a general quantum 
system with underlying chaotic classical dynamics has statistical properties of
energy levels described by random matrices theory has been confirmed by 
theoretical studies of numerous chaotic systems, see e.g. 
\cite{haake90,bohigas91}. 
Experimental verifications on the other hand are
much less abundant \cite{zimmermann88,walther98}
and, as far as we know, concern only systems possessing
anti-unitary (generalized time-reversal) symmetry where the level statistics
can be modeled by random matrices of the Gaussian orthogonal ensemble (GOE)
\cite{mehta91}. 
Realization of a quantum system with quadratic level repulsion as is
typical for matrices from the Gaussian unitary ensemble (GUE) \cite{mehta91}
requires breaking
of any anti-unitary symmetry invariance. E.g. for atomic or
molecular systems it means one would have to apply magnetic field  
inhomogeneous across the molecule  
being well experimentally controlled on
such a small scale \cite{haake90}. 
Such a requirement is rather unattainable even in the
Rydberg regime of excitation. While a real quantum system with GUE statistics 
has not been realized experimentally, there are experiments with microwave
cavities (so-called {\it wave chaos} experiments) 
where anti-unitary symmetry can be broken by applying some ferrite 
devices \cite{so95,stoffregen95}. 

Recently, however, it has been shown \cite{sacha99b,sacha00}
that to break anti-unitary symmetry
invariance, in atomic systems, it is not necessary to employ magnetic field.
Indeed a combination of elliptically polarized microwaves
with a static electric field applied to, e.g., hydrogen (H) atoms can do 
this work as well.
In the previous studies \cite{sacha99b,sacha00}
we have restricted ourselves to a
case when the microwave field 
is resonant with the classical motion of the electron, i.e. the ratio of
microwave frequency, $\omega$, to Kepler frequency, $\omega_K=1/n_0^3$ (where
$n_0$ is the principal quantum number of H atom), is an integer number.
It has been found that such a system can reflect level statistics very 
close to the GUE prediction.

In this paper we study a more general situation of non-resonant driving of
the atoms and show that appropriate choice of the system parameters allows us 
to reach statistical properties expected for matrices of the GUE. 
We consider weak fields limit where 
quantum results may be obtained by the lowest orders of the perturbation theory.
Classical investigation of the system
behavior is carried out also in terms of the perturbation calculations. It 
allows us to find out for which system parameters the secular motion of the
electronic ellipse reveals chaotic dynamics.

The Hamiltonian of a realistic three-dimensional H atom placed in a
static
electric field and driven by an elliptically polarized microwave field 
in atomic units, neglecting relativistic effects, assuming
infinite mass of the nucleus, and employing dipole approximation reads:
\be
H=\frac{\mathbf p^2}{2}-\frac{1}{r}+F(x\cos\omega t+\alpha y\sin\omega
t)+
{\mathbf E}\cdot {\mathbf r},
\label{h}
\ee
where $F$, $\alpha$ and $\omega$ are, respectively,
the amplitude, degree of elliptical polarization and frequency of 
the microwave field while ${\mathbf E}$ stands for the static electric
field vector. As the external perturbation is time periodic, one may apply
Floquet formalism and look for quasienergy levels of the system. 

We assume very 
small field amplitudes which allows including only the 
lowest non-vanishing terms in the quantum 
effective Hamiltonian \cite{cct92} describing 
the dynamics of a given $n_0$ hydrogenic
manifold. The static electric field contributes already in the first order 
of $E$ as the states with fixed $n_0$ are coupled among each other by the 
${\mathbf E}\cdot {\mathbf r}$ operator.
On the other hand there is no {\it
direct} coupling between the states due to the microwave perturbation. 
Indeed the states can be coupled only {\it indirectly} by the 
process of absorption
and emission of microwave photons. 
So that the first non-vanishing term is second order in $F$.
The final matrix (with $n_0^2$ dimension) of the quantum effective 
Hamiltonian, i.e. the Hamiltonian in the first order in $E$ and 
second order in $F$, 
is then diagonalized by standard routines.
The details of the quantum perturbation calculations can be found in 
\cite{sacha00}, here we would only like to stress that
the whole physically realistic 
problem is reduced to the analysis of the finite $n_0^2$ dimensional
Hilbert space where $1/n_0$ plays the r$\hat{\mbox{o}}$le of the effective 
Planck constant. 

We are now turning to classical analysis of the system. 
The range of our interests is 
the high frequency regime, i.e. $\omega >\omega_K$, which means that 
for weak external
fields we have two fast degrees of freedom in the system, 
i.e. the position of the
electron on an elliptical orbit and the phase of the microwave field, and two
slow degrees of freedom corresponding to the orientation of 
the elliptical trajectory. One
can get rid of the fast degrees of freedom by means of the classical
perturbation theory \cite{lichtenberg83}
which results in the classical effective Hamiltonian
describing slow precession of an electronic ellipse.
The perturbation calculation can be easily carried out employing the Lie 
method \cite{lichtenberg83}, 
actually it closely follows the similar procedure applied in 
\cite{abdd97} to the H atom perturbed by linearly polarized microwave field.
The first stage is to express the Hamiltonian (\ref{h}) in terms of the 
action-angle
variables of the unperturbed hydrogen atom. The new pairs of the canonically
conjugate variables are $(J,\Theta)$, i.e. principal action (analog of the
principal quantum number, $n_0$) and position of the electron on an ellipse
respectively; $(L,\Psi)$, i.e. angular momentum of the electron and conjugate
angle; and finally $(M,\Phi)$, i.e. angular momentum projection on $z$ axis 
and angle of rotation around this axis \cite{sacha98c}.
Averaging the resulting Hamiltonian over $\Theta$ and $t$ immediately gives the 
first order contribution to the classical effective Hamiltonian
\bea
H^{(1)}(L,\Psi,M,\Phi)&=&-\frac{3}{2}n_0^2\sqrt{1-\frac{L^2}{n_0^2}}\cr
 && \left[E_x
          \left(\cos\Phi\cos\Psi-
\frac{M}{L}\sin\Phi\sin\Psi\right)\right. \cr
&&  +E_y\left(\sin\Phi\cos\Psi+\frac{M}{L}
\cos\Phi\sin\Psi\right) \cr
&& \left. +E_z\sqrt{1-\frac{M^2}{L^2}}\sin\Psi
\right].
\label{h1}
\eea
The second order contribution of microwave field requires
calculating the generating function, $w$ \cite{lichtenberg83}
which is the solution of the following equation
\be
\frac{\partial w}{\partial t}+\omega_K\frac{\partial w}{\partial \Theta}=
-H_{micro},
\label{gen}
\ee
where $H_{micro}$ is the microwave part 
of the Hamiltonian (\ref{h}) expressed in
the action-angle variables (explicit expression for $H_{micro}$ as a Fourier
series can be found in \cite{sacha98c}). 
The solution for $w$ is given as an infinite series
with terms containing $1/(\omega\pm m\omega_K)$, where $m$ is an integer number.
For resonant driving, i.e. $\omega/\omega_K\approx m$, one faces 
the small
denominators problem \cite{lichtenberg83} but here we are not affected by 
this problem as we are interested in a
non-resonant perturbation. Having calculated 
$w$ it is straightforward task to obtain
the second order contribution to the effective Hamiltonian by averaging the 
Poisson bracket of $w$ and $H_{micro}$ over $\Theta$ and $t$, i.e.
\be
H^{(2)}(L,\Psi,M,\Phi)=\frac{1}{2}\langle \{w,H_{micro}\}\rangle_{\Theta,t},
\label{second}
\ee
(we omit 
the lengthy explicit formula for $H^{(2)}$ here). 
The final classical effective Hamiltonian reads
\be
H^{eff}(L,\Psi,M,\Phi)=-\frac{1}{2n_0^2}+H^{(1)}+H^{(2)}.
\label{final}
\ee
This is the classical counterpart of the quantum effective Hamiltonian, namely
first order in the static electric field and second order in microwave field.
The classical Hamiltonian (\ref{final}) possesses scaling symmetry, i.e. 
one can get rid of one of the parameters of the system. Introducing 
$F_0=n_0^4F$, $E_0=n_0^4E$, $\omega_0=n_0^3\omega$, $L_0=L/n_0$, $M_0=M/n_0$ 
and $H^{eff}_0=n_0^2H^{eff}$ the dynamics becomes
independent of the particular choice of the 
$n_0$ hydrogenic manifold. 

For a general fields 
configuration the secular motion in the $(L,\Psi,M,\Phi)$ phase space
is not integrable and to investigate classical
dynamics we have to perform numerical integration of the equations of motion
generated by the Hamiltonian (\ref{final}).
For the linear microwave polarization without additional static electric field
considered in \cite{abdd97} the secular motion was one dimensional and
employing the WKB quantization rule \cite{haake90}
the authors were able to get semiclassical 
predictions for quasienergy levels in a very good agreement with exact quantum
numerical data.

\begin{figure}
\epsfig{file=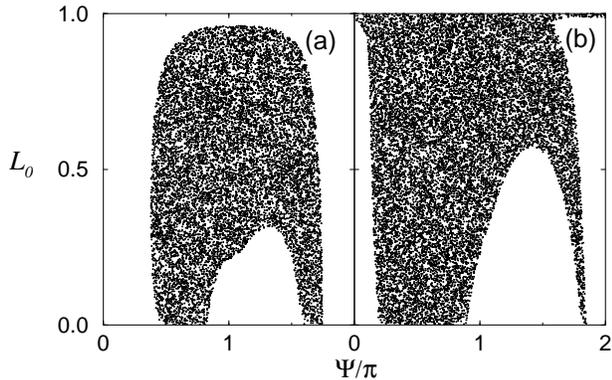%
,scale=0.34,angle=-90
}
\caption{Poincar\'e surface of section (at $\Phi=0$) 
of the classical secular motion, Eq.~(\protect{\ref{final}}), for the 
hydrogen atom placed 
in a static electric field, with the amplitude $E_0=0.00028$, 
and driven by an elliptically
polarized ($\alpha=0.4$) microwave field with frequency $\omega_0=1.304$
and amplitude $F_0=0.02$. The coordinates used for the plot are the scaled
angular momentum $L_0=L/n_0$ and its canonically conjugate angle $\Psi$.
Panel (a) corresponds to the 
anti-unitary invariant case, i.e. the orientation of
static field vector is $\phi=0$, $\psi=\pi/4$; with the energy 
$H_0^{eff}=-0.50008$.
Panel (b) is related to the broken anti-unitary invariance, i.e.
$\phi=0.3\pi$, $\psi=\pi/4$; with the energy $H_0^{eff}=-0.5002$.
Note that, for the parameters chosen,
not the whole $(L_0,\psi)$ space is accessible.}
\label{one}
\end{figure}

For elliptically polarized microwaves and general orientation of the
static field vector
the anti-unitary symmetry invariance is broken. Only when either
$E_x=0$ or $E_y=0$ the system is invariant with respect to the time-reversal
combined with $x\rightarrow -x$ or $y\rightarrow -y$ transformations
respectively, see (\ref{h}). As an example of a general elliptical
polarization case we have further on analyzed
the degree of the polarization 
$\alpha=0.4$. The microwave frequency has been chosen as 
$\omega_0=1.304$ and the amplitude as $F_0=0.02$ which is well in the range
where, for the linear microwave polarization, the WKB calculations give good
agreement with the exact numerical results \cite{abdd97}.
Then by investigating Poincar\'e surface of 
section we have found that 
the amplitude $E_0=0.00028$ and the orientation of the static field vector
$\phi\approx 0.3\pi$, $\theta\approx\pi/4$ (where $\theta$, $\phi$ are usual
spherical angles) correspond to chaotic dynamics in the energy
interval $H_0^{eff}\in [-0.50032, -0.50008]$. Putting $\phi=0$ ($E_y=0$) one
recovers anti-unitary symmetry of the system, then the classical dynamics is 
found to be predominately chaotic for $H_0^{eff}\in [-0.50016, -0.49992]$.
Fig.~\ref{one} shows examples of the phase space structures for both 
$\phi=0$ and $\phi=0.3\pi$. 

\begin{figure}
\epsfig{file=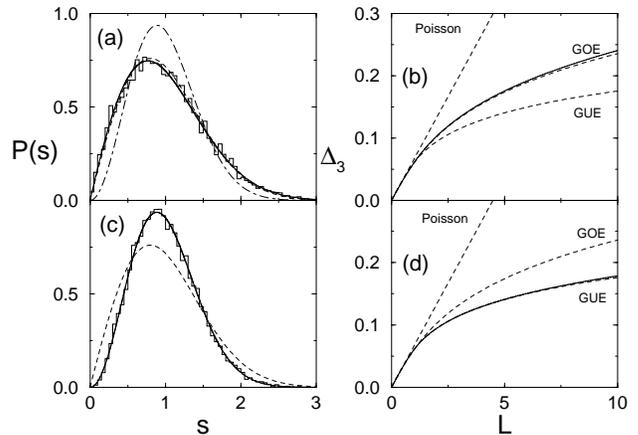%
,scale=0.34,angle=-90
}
\caption{ Nearest neighbor spacing distribution and spectral rigidity,
$\Delta_3$ statistics, for hydrogen atom placed in a static electric field
and illuminated by microwave field, for the case with [panels (a)-(b)]
and without [panels (c)-(d)] anti-unitary 
symmetry invariance, compared with predictions of random matrices ensembles. 
The data have been collected for $n_0=50\div 59$ with the same
fields parameters as in Fig.~\protect{\ref{one}}.  
In panels (a) and (c) solid lines indicate the best fitting 
Izrailev distributions, while dashed and dash-dotted lines 
correspond to GOE and GUE distributions respectively
(in panel (c) the dash-dotted line is invisible behind the solid one). 
Panels (b) and (d): solid and dotted (hardly
visible behind the solid lines) lines correspond to numerical 
data and their best fits, while dashed lines indicate Poisson, GOE and
GUE predictions as indicated in the figure. 
}
\label{two}
\end{figure}

Having done classical analysis we can switch to quantum 
calculation results. 
In order to get reasonable statistics for quasienergy levels we have
diagonalized the matrix of the quantum effective Hamiltonian for 
different $n_0$ manifolds in the range $n_0=50\div 59$. From each
diagonalization we have separated and unfolded spectrum in the energy 
intervals corresponding to chaotic classical dynamics. 
This procedure has been applied to the anti-unitary invariant case ($\phi=0$)
and to the case with broken anti-unitary symmetry ($\phi=0.3\pi$).
Fig.~\ref{two} shows histograms of the nearest neighbor spacing (NNS) 
distributions of the quasienergy levels (there are about $10^4$ spacings in each
of the data sets) together with the plots of the Wigner surmises 
\cite{bohigas91} for the GOE and GUE. The figure also
shows the spectral rigidities, i.e.
$\Delta_3$ statistics \cite{bohigas91}. 
The qualitative agreement of the data with the random 
matrices theory is apparent, especially for the case with broken 
anti-unitary symmetry. 

To focus on quantitative measure we have fitted
theoretical NNS distributions to the data (to avoid the dependence of the
results on bin size the distributions have been fitted to the cumulative
histograms \cite{prozen93}). 
For the anti-unitary invariant case the best fitting Berry-Robnik
distribution (i.e. the distribution 
for independent superposition of Poisson and GOE spectra \cite{berry84})  
results in the parameter value (relative measure of the chaotic 
part of the phase
space) $q=0.98$ with $\chi^2/N=0.3$, i.e. chi-squares divided by the 
number of spacings $N$.
Employing the Izrailev distribution \cite{izrailev90,casati91}
we get the value of the levels repulsion
parameter $\beta=0.96$ with $\chi^2/N=0.8$. For the spectral rigidity the best
fitting
$\Delta_3$ statistics corresponding to an independent superposition of Poisson 
and GOE spectra \cite{bohigas91}
results in $q=0.99$ which is in agreement with the value obtained from the
Berry-Robnik distribution fit.

In the case with broken anti-unitary symmetry one gets the following 
parameters values: Berry-Robnik statistics (now it corresponds to independent
superposition of the Poisson and GUE spectra
\cite{robnik99}) $q=1$ with $\chi^2/N=0.2$,
Izrailev distribution $\beta=2.05$ with $\chi^2/N=0.3$ and from the spectral
rigidity fit $q=1$.

The presented results show undoubtedly that the system under consideration 
can reflect the Bohigas, Giannoni and Schmit conjecture both with 
or without anti-unitary symmetry. The question is if such a behavior
can be observed experimentally? 
With this respect let 
us consider the case with broken generalized
time-reversal symmetry. As there is no discrete symmetry in the system 
one has no problems with separation of overlapping spectra. On the other hand it
results in a big density of states and requires high experimental resolution.
We have presented the data for hydrogenic manifolds in the range 
$n_0=50\div 59$
in order to have good statistics but our results for $n_0=40\div 49$ reveal
the very same behavior. For example for $n_0=40$ and $F=40$~V/cm,
$\omega=2\pi\cdot 134$~GHz, $E=0.56$~V/cm the average level 
spacing is of order $10^{-4}$~cm$^{-1}$ and a measurement 
in the energy range between $-68.628$~cm$^{-1}$ and $-68.595$~cm$^{-1}$ 
provides about 500 levels. It sounds feasible
experimentally.

The previous studies \cite{sacha99b,sacha00}
have been devoted to resonant microwave driving of H atoms
placed in a static electric field. The present work deals with a general
non-resonant driving and shows that intra-manifold chaotic dynamics
\cite{uzer94,uzer97,main98,sacha99b,sacha00}, i.e. the
situation when states corresponding to a fixed principal quantum number $n_0$
are mixed significantly only among each other and underlying classical 
dynamics is
irregular, is not restricted to a particular resonant case but exists widely 
in the frequency domain. The presented behavior is 
field amplitudes independent,
i.e. if $F\rightarrow 0$ and $E\rightarrow 0$ but $E/F^2=const$ the structure
of the phase space corresponding to secular motion remains unchanged as 
are the
statistical properties of quasienergy levels. This is a signature of
inapplicability of the Kolmogorov-Arnold-Moser theorem \cite{lichtenberg83}
to a highly degenerate Coulomb problem. 
For high field amplitudes when hydrogenic manifolds can not be
considered isolated and the inter-manifold mixing comes to the scene
our perturbation approach, obviously, becomes irrelevant. However 
this regime, for a realistic three-dimensional problem, 
still constitutes a challenge for the theory.

The author is grateful to Dominique Delande and
Jakub Zakrzewski for discussions.
The results were attained with the assistance of the Alexander von Humboldt
Foundation. The partial support of KBN under project 2P302B00915 is 
acknowledged.

\end{multicols}

\end{document}